\documentclass[aps,twocolumn]{revtex4}
\usepackage{epsf}
\begin{document}

\title{Electron pair current through the correlated  quantum dot}

\author{T. Doma\'nski\footnote{Corresponding author: 
       {\sf doman@kft.umcs.lublin.pl}} 
       and A. Donabidowicz}
\affiliation{
           Institute of Physics, M.\ Curie Sk\l odowska University, 
           20-031 Lublin, Poland 
}
\date{\today}

\begin{abstract}
We study the charge current transmitted through the correlated 
quantum dot characterized by a finite magnitude of the Coulomb 
interaction $|U|$. At low temperatures the correlations can lead 
to a formation of the spin (for $U\!>\!0$) or charge (for $U\!<\!0$) 
Kondo states which qualitatively affect the transport properties. 
We explore the influence of charge Kondo effect on the electron 
pair tunneling introducing the auxiliary two-component model 
which describes fluctuations between the empty and doubly 
occupied states on QD.
\end{abstract}

\maketitle

\section{Introduction}
Electronic transport through the correlated quantum dots 
(QDs) has recently attracted a considerable interest both 
from a point of view of fundamental research as well as 
practical applications \cite{books,book2}. Both these aspects 
are possible because of a large degree of flexibility for 
adjusting a coupling of QD to the external leads and 
a controlled positioning of the quantized QD levels 
by the gate voltage.

The tunability of the QD systems enabled controlled 
exploration of the Kondo effect \cite{Goldhaber-98}, 
a phenomenon first observed in metals with diluted 
magnetic impurities \cite{Hewson}.
At sufficiently low temperatures $T\!<\!T_{K}$ (where 
$T_{K}$ denotes the, so called, Kondo temperature) the 
spin of QD combines into a singlet state with the spins 
of itinerant electrons from the leads. In consequence the QD 
spectrum develops a narrow resonance at the chemical potential
which at low temperatures enhances the zero-bias conductance 
to the perfect unitary limit value $2e^{2}/h$ \cite{Goldhaber-98}. 
On a microscopic level the underlying Kondo physics arises 
from an effective antiferromagnetic interaction between 
the QD and mobile electrons as envisaged by Schrieffer 
and Wolf within the perturbative canonical transformation 
\cite{Schrieffer-66}. 

Recently several authors \cite{negativeU,Koch} have pointed out 
that molecular quantum dots affected by the bosonic degrees of 
freedom such as e.g.\ phonons could show up (besides the 
side-band structure) signatures of the charge Kondo effect. 
This phenomenon has been previously theoretically predicted 
for the heavy fermion compounds \cite{Coleman-91} and it might 
occur when the bipolaronic shift lowers the charging energy to 
a negative value $U\!<\!0$. The essential physics involved in 
the charge Kondo effect relies on a neutralization of the electron 
pair charge at the negative $U$ center (in the present context 
on QD) by electrons (or holes) from the adjacent leads. Due to a 
preferred double occupancy of QD there is activated a mechanism 
of the electron pair tunneling which manifests the qualitative 
features of charge Kondo effect in the differential conductance 
\cite{Koch,Hwang-07} and in the thermopower \cite{Wysokinski-08}.  

In what follows we propose a phenomenological two-component model 
which allows for a simple description of charge tunneling under 
the circumstances when the empty $\left| 0 \right>$ and doubly 
occupied $\left| \uparrow\downarrow \right>$ states are 
degenerate (which is a necesarry condition for realization of 
the charge Kondo effect \cite{negativeU,Koch,Coleman-91}). We 
discuss some preliminary results for the QD spectral function 
and the differential conductance obtained in the symmetric case 
using approximate treatment for the on-dot correlations.

\section{spin versus charge Kondo effects}

For a description of charge tunneling through the %single level 
correlated quantum dot we use the Anderson model \cite{Hewson}
\begin{eqnarray}
\hat{H} & = & \sum_{{\bf k},\beta,\sigma } \xi_{{\bf k}\beta}
\hat{c}^{\dagger}_{{\bf k} \beta \sigma} \hat{c}_{{\bf k} \beta 
\sigma} +  \hat{H}_{QD} 
\label{model} \\ 
&&+ \sum_{{\bf k},\beta, \sigma} \left( V_{{\bf k} \beta} 
\hat{d}_{\sigma}^{\dagger} \hat{c}_{{\bf k} \beta \sigma} + 
V_{{\bf k} \beta}^{*} \hat{c}^{\dagger}_{{\bf k} \beta \sigma}
\hat{d}_{\sigma} \right) , \nonumber \\
\hat{H}_{QD} & = & \sum_{\sigma} \varepsilon_{d} \; 
\hat{d}^{\dagger}_{\sigma} \hat{d}_{\sigma} + U \; 
\hat{d}_{\uparrow}^{\dagger} \hat{d}_{\uparrow} \; 
\hat{d}_{\downarrow}^{\dagger} \hat{d}_{\downarrow} .
\label{QD_hamil}
\end{eqnarray}
Operators $\hat{c}_{{\bf k} \beta \sigma}^{(\dagger)}$ correspond 
to annihilation (creation) of electrons in the left $\beta\!=\!L$ 
or right h.s.\ $\beta\!=\!R$ lead. The energies $\xi_{{\bf k} \beta 
\sigma}\!=\!\epsilon_{{\bf k} \sigma}\!-\!\mu_{\beta}$ are measured 
with respect to the chemical potentials which under nonequilibrium 
conditions are shifted by an applied bias $V$ through $\mu_{L}
\!-\!\mu_{R}\!=\!eV$. The other terms containing $V_{{\bf k}\beta}^{(*)}$ 
describe hybridization of the QD to external leads. As usually  
$\hat{d}_{\sigma}^{(\dagger)}$ denote the annihilation 
(creation) of electron at the QD energy level $\varepsilon_{d}$ 
and $U$ is the on-dot Coulomb potential. 

Let us first start by focusing on a widely studied case of the repulsive 
charging energy $U\!>\!0$. Conditions necesarry for a appearance of 
the Kondo effect can be explained using a perturbative treatment for 
the hybridization term $\hat{H}_{hyb}=\sum_{{\bf k},\beta, \sigma} 
[ V_{{\bf k} \beta} \hat{d}_{\sigma}^{\dagger} \hat{c}_{{\bf k} 
\beta \sigma}  + h.c.]$. The unitary transformation $\hat{\tilde{H}}
=e^{\hat{A}} \hat{H} e^{-\hat{A}}$ with the antihermitean generating 
operator $\hat{A}=\hat{\cal{A}}-\hat{\cal{A}}^{\dagger}$ where $\hat{\cal{A}} 
= \sum_{{\bf k},\beta,\sigma} \frac{V_{{\bf k} \beta}}{\varepsilon_{d}
-\xi_{{\bf k}\beta}} \left[ \frac{U}{\varepsilon_{d} + U - \xi_{{\bf k}
\beta}}  \hat{d}_{-\sigma}^{\dagger} \hat{d}_{-\sigma}  - 1 \right] 
\hat{c}_{{\bf k}\beta\sigma}^{\dagger} \hat{d}_{\sigma}$ eliminates 
$\hat{H}_{hyb}$ up to the quadratic terms \cite{Schrieffer-66}. Since 
for $U>0$ the double occupancy of QD is energetically expensive one can 
restrict to the subspace of singly occupied states when the transformed 
Hamiltonian reduces to the spin Kondo model \cite{Schrieffer-66}
\begin{eqnarray}
\hat{\tilde{H}}_{spin}^{Kondo} = \sum_{{\bf k},\beta,\sigma } 
\xi_{{\bf k}\beta} \hat{c}^{+}_{{\bf k} \beta \sigma} 
\hat{c}_{{\bf k} \beta \sigma}  \! -\!\!\!\!\! \sum_{{\bf k},{\bf q},
\beta,\beta'} J_{{\bf k},{\bf q}}^{\beta,\beta'} \;\; 
\hat{\vec{S}}_{d} \cdot \hat{\vec{S}}_{{\bf k}\beta,{\bf q}\beta'}
%( \hat{c}_{{\bf k}\beta\sigma}^{\dagger} 
%\vec{\sigma}_{\sigma,\sigma'} \hat{c}_{{\bf q}\beta'\sigma'}) 
. \nonumber \\ \label{SW}
\end{eqnarray}
The QD spin operator $\hat{\vec{S}}_{d}$ can be conveniently expressed 
through $\hat{S}_{d}^{+}=\hat{d}_{\uparrow}^{\dagger}\hat{d}_{\downarrow}$, 
$\hat{S}_{d}^{-}=\hat{d}_{\downarrow}^{\dagger} \hat{d}_{\uparrow}$, 
and $\hat{S}_{d}^{z}=\frac{1}{2}(\hat{d}_{\uparrow}^{\dagger} \hat{d}
_{\uparrow} - \hat{d}_{\downarrow}^{\dagger}\hat{d}_{\downarrow})$
and similarly $\hat{S}_{{\bf k}\beta,{\bf q}\beta'}^{+}=\hat{c}_{
{\bf k}\beta \uparrow}^{\dagger}\hat{c}_{{\bf q} \beta \downarrow}$
etc. Near the Fermi surface the effective coupling $J_{{\bf k},{\bf q}}
^{\beta,\beta'}$ simplifies to \cite{Schrieffer-66} $J_{{\bf k}_{F},
{\bf k}_{F}}^{\beta,\beta'} = \frac{U}{\varepsilon_{d} (\varepsilon_{d}
+U)} \; V_{{\bf k}_{F}\beta} V^{*}_{{\bf k}_{F} \beta'}$. In the 
regime of antiferromagnetic coupling $J_{{\bf k}_{F},{\bf k}_{F}}
^{\beta,\beta'}\!<\!0$  and at sufficiently low temperatures 
$T<T_{K}$ the magnetic moment of QD is perfectly screened by 
spins of the itinerant electrons. For typical QDs the value 
of Kondo temperature $T_{K} \simeq \frac{\sqrt{U\Gamma}}{2} 
\mbox{exp}\left\{\frac{\pi \varepsilon_{d} \left( \varepsilon_{d}
\!+\!U\right)}{U\Gamma } \right\}$ (where $\Gamma \simeq 2 \pi 
\sum_{\beta} |V_{{\bf k}_{F}\beta}|^{2} \rho_{\beta}(\varepsilon_{F})$
and $\rho_{\beta}(\varepsilon_{F})$ is the density of states at 
the Fermi level) is of the order of hundreds mK. Appearance of 
such Kondo resonance pinned at the electrodes' chemical potentials 
enhances the zero bias conductance and this behavior has been 
indeed observed experimentally \cite{Goldhaber-98}.

Formally for the negative $U$ model the canonical transformation 
can be done in the same way. Since the empty and doubly occupied 
QD states are then energetically more favorable therefore one 
focuses mainly on the terms describing pair tunneling  
$\sum_{{\bf k},{\bf q},\beta,\sigma,\sigma'} \left( 
J_{{\bf k},{\bf q}}^{\beta,\beta'}\; \hat{d}_{\sigma}^{\dagger}
\hat{d}_{-\sigma}^{\dagger} \hat{c}_{{\bf k}\beta-\sigma'} 
\hat{c}_{{\bf q} \beta' \sigma'} + \mbox{h.c.} \right)$ \cite{Koch}.
Taraphder and Coleman \cite{Coleman-91} have shown that in the 
symmetric case $\varepsilon_{d}+U/2\!=\!0$ (here assuming also
$V\!=\!0$) the attractive $U<0$ model becomes exactly isomorphic 
to its repulsive $U>0$ counterpart (\ref{model}) under the following 
particle-hole (p-h) transformation $\hat{d}_{\downarrow}^{\dagger} 
\rightarrow - \hat{d}_{-\downarrow}$, $\hat{c}_{{\bf k}\beta 
\downarrow}^{\dagger} \rightarrow \hat{c}_{-{\bf k}\beta 
\downarrow}$. Outside the symmetric situation the p-h 
transformation still renders the structure of (\ref{model}) 
with an additional Zeeman field $B^{z}\!=\!2\varepsilon_{d}+U$
(see Ref.\ \cite{Koch} for details). The effective Hamiltonian 
\cite{Schrieffer-66} operating on the relevant empty and doubly 
occupied QD states can be then expressed by  \cite{Coleman-91}
\begin{eqnarray}
\hat{\tilde{H}}_{charge}^{Kondo} &=& \sum_{{\bf k},\beta,\sigma } 
\xi_{{\bf k}\beta} \hat{c}^{+}_{{\bf k} \beta \sigma} 
\hat{c}_{{\bf k} \beta \sigma}  \label{chargeKondo} \\
&+& 2\!\! \sum_{{\bf k},{\bf q},
\beta,\beta'} J_{{\bf k},{\bf q}}^{\beta,\beta'} \;\; 
\hat{\vec{{\cal{T}}}}_{d} \cdot 
\hat{\vec{{\cal{T}}}}_{{\bf k}\beta,{\bf q}\beta'} 
+ \hat{\cal{T}}_{d}^{z} \; B^{z}
\nonumber
\end{eqnarray}
where $\hat{\cal{T}}_{d}^{+}=\hat{d}_{\uparrow}^{\dagger} 
\hat{d}_{\downarrow}^{\dagger}$, $\hat{\cal{T}}_{d}^{-}=
\hat{d}_{\downarrow} \hat{d}_{\uparrow}$, and $\hat{\cal{T}}
_{d}^{z}=\frac{1}{2}(\hat{d}_{\uparrow}^{\dagger} \hat{d}
_{\uparrow} + \hat{d}_{\downarrow}^{\dagger}\hat{d}
_{\downarrow} -1)$. Using this pseudospin representation 
the spin Kondo effect can be directly translated into the 
charge Kondo effect of the model (\ref{chargeKondo}). 
The latter takes place in a vicinity of the degeneracy 
point $\varepsilon_{d}+U/2 \sim 0$. So far some of its 
signatures in the pair tunneling through the QDs have 
been partly examined in Refs \cite{negativeU,Koch}.

\section{The two-component model}

To consider the situation when the empty and doubly 
occupied states of QD are degenerate we introduce 
the following auxiliary model 
\begin{eqnarray}
\hat{H}_{QD} &=& \sum_{\sigma} E_{d} \; 
\hat{d}_{\sigma}^{\dagger} \hat{d}_{\sigma} + g \left( 
\hat{b}^{\dagger} \; \hat{d}_{\downarrow} \hat{d}_{\uparrow} 
+  \hat{d}_{\uparrow}^{\dagger} \hat{d}_{\downarrow}^{\dagger} 
\; \hat{b} \right) \nonumber \\ & & 
+ E_{pair} \; \hat{b}^{\dagger} \hat{b} .
\label{BFM}
\end{eqnarray}
Whenever the electron pair happens to arrive on the d-QD  
we let it be stored on  the side-coupled buffer described by 
the operators $\hat{b}^{(\dagger)}$. They obey the hard-core
boson relations \cite{Ranninger-84} so that at most only 
one electron pair can be allocated on this side-coupled b-QD.  
Similar lattice version of this model has been proposed in the 
solid state physics to account for the bipolaron superfluidity 
in the crossover between the adiabatic and antiadiabatic 
regimes \cite{Ranninger-84}.

\begin{figure}
\epsfxsize=9cm
\centerline{\epsffile{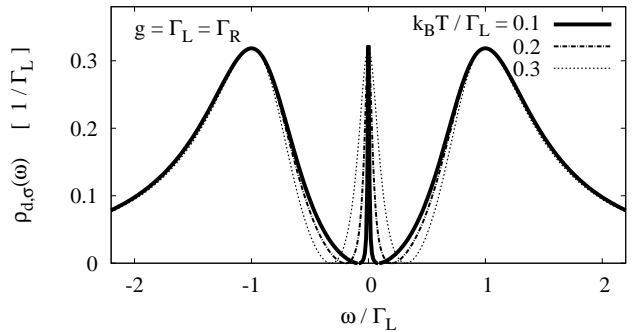}}
\caption{Spectral function of the molecular quantum dot for
the equilibrium situation ($V\!=\!0$) in the symmetric case 
$E_{pair}\!=\!0$, $E_d\!=\!0$ for several temperatures. We 
used $g\!=\!\Gamma_{L}\!=\!\Gamma_{R}$ assuming the wide-band 
limit $\Gamma_{\beta}=0.001D$, where $D$ is the bandwidth of
conduction electrons.}
\label{density}
\end{figure}

Loosely speaking the correspondence between (\ref{BFM}) and 
the negative $U\!<\!0$ Hamiltonian (\ref{QD_hamil}) holds 
via substitutions $g=U$, $E_{pair}=2\varepsilon_{d}+U$. This 
relation can be further supported analytically in the Lagrangian 
language after performing the Hubbard-Stratonovich transformation
which eliminates $U \hat{d}_{\uparrow}^{\dagger} \hat{d}_{\uparrow} 
\hat{d}_{\downarrow}^{\dagger} \hat{d}_{\downarrow}$ through  
the additional bosonic fields, here denoted by $\hat{b}^{(\dagger)}$.

We will show that the two-component QD described by (\ref{BFM}) 
is able to capture the charge Kondo effect \cite{Coleman-91} 
known for the negative $U$ Anderson model \cite{Anderson} near 
the symmetric situation. The underlying mechanism is driven
by suppressing the quantum fluctuations between the empty and doubly 
occupied states. Within the phenomenological model (\ref{BFM}) 
such fluctuations are present even in the case of isolated molecular 
quantum dot i.e.\ for $V_{{\bf k}\beta}\!=\!0$. Out of 8 possible 
configurations being a product of the fermionic states $|0>$, 
$|\uparrow>$, $|\downarrow>$, $|\uparrow\downarrow>$ and the 
hard-core bosonic ones $|0)$, $|1)$  two of them $\uparrow
\downarrow> \otimes |0)$ and $|0> \otimes |1)$ get mixed due 
to the Andreev-like interaction. A complete set of the 
eigenstates can be obtained using the transformation
\cite{Domanski-BFM}
\begin{eqnarray}
|B> & = &      \sin(\varphi) \; |0> \otimes |1) \hspace{2mm} 
+ \hspace{2mm} \cos(\varphi) \; |\uparrow\downarrow> 
\otimes |0) ,
\\
|A> & = &      \cos(\varphi) \; |0> \otimes |1) \hspace{2mm}
- \hspace{2mm} \sin(\varphi) \; |\uparrow\downarrow>
\otimes |0) 
\end{eqnarray}
with $\tan(2\varphi) = \frac{2g}{2E_{d} - E_{pair}}$.
For such limit $V_{{\bf k}\beta}\!\rightarrow\!0$ the Green's function 
${\cal{G}}_{d}(\omega) = \langle \langle \hat{d}_{\sigma} ; \hat{d}
_{\sigma}^{\dagger} \rangle \rangle_{\omega}$ acquires the three pole 
structure \cite{Domanski-BFM}
\begin{equation}
{\cal{G}}^{V_{{\bf k}\beta}=0}_{d}(\omega) = \frac{\cal{Z}}
{ \omega - E_{d} } + \left( 1 - {\cal{Z}} \right) 
\left[ \frac{u^{2}}{\omega - E_{+}} +
\frac{v^{2}}{\omega - E_{-}} \right]
%\equiv \frac{1}{\omega - \varepsilon_{d} - 
%\Sigma_{d}^{V_{{\bf k}\beta}=0}(\omega) }
\label{GF_at}
\end{equation}
where 
\begin{eqnarray}
v^{2},u^{2} & = &\frac{1}{2} \; \left( 1 \mp
\frac{1}{\gamma} \right) ,
\\
E_{-},E_{+} & = & \frac{1}{2} \left[ E_{pair} 
\mp (2 E_{d} - E_{pair}) \gamma \right] 
\end{eqnarray}
with $\gamma^{2} = 1+4g^{2}/(2E_{d}-E_{pair})^{2}$ and 
${\cal{Z}} = \left( 1 + e^{-E_{d}/k_{B}T} + e^{-(E_{d}+E_{pair})
/k_{B}T} + e^{-(2E_{d} + E_{pair})/k_{B}T} \right)$ 
$/\left[ 1 + 2e^{-E_{d}/k_{B}T} + e^{-(E_{d}+E_{-})/k_{B}T} 
\;+\; e^{-(E_{d}+E_{+})/k_{B}T} \;+\right.$ \newline
$\left. 2e^{-(E_{d}+E_{pair})/k_{B}T} + 
e^{-(2E_{d}+E_{pair})/k_{B}T}\right]$.
The spectral weight ${\cal{Z}}$ of single particle level is
gradually depleted for a decreasing temperature and its amount
is transferred to the bonding $E_{-}$ and antibonding $E_{+}$ 
levels.

To get some insight into the many-body physics we employ 
a simple approximation based on the following substitution 
for the self-energy (see the discusion in Section 12.5 of 
the review book \cite{book2})
\begin{eqnarray}
{\cal{G}}_{d}(\omega)^{-1} = {\cal{G}}^{V_{{\bf k}\beta}
=0}_{d}(\omega)^{-1} - \sum_{{\bf k}, \beta} \frac{
|V_{{\bf k}\beta}|^{2}}{\omega-\varepsilon_{{\bf k}\beta}}. 
\label{GF} 
\end{eqnarray}
We shall focus on the wide-band limit when the hybridization 
couplings $\Gamma _{\beta}(\omega)\!=\!2\pi\sum_{\bf k} |V_{
{\bf k}\beta}|^{2}\delta\left(\omega-\varepsilon_{{\bf k}\beta}
\right)$ can be treated as constant quantities $\Gamma_{\beta}$. 
Figure \ref{density} shows the spectral function $\rho_{d,\sigma}
(\omega)=-\frac{1}{\pi} \mbox{Im}{\cal{G}}_{d}(\omega+i0^{+})$ 
obtained for the symmetric case $E_{pair} \!=\!0$, $E_{d}\!=\!0$ 
when $u^2\!=\!0.5\!=\!v^{2}$ and $E_{+}\!=\!g$, $E_{-}\!=\!-g$. 
We notice the three-peak structure where the middle one is 
sensitive to temperature due to transfer of the spectral 
weight ${\cal{Z}}$. Let us emphasize that this behavior 
is typical for the Dicke effect known in quantum optics 
where the narrow/broad energy features correspond to 
the states weakly/strongly coupled to the electromagnetic 
field and they contribute the subradiant/superradiant 
emission lines \cite{Brandes-04}. Recently a similar 
concept of the Kondo-Dicke effect has been proposed 
in mesoscopic physics for a set of three vertically 
coupled quantum dots \cite{Trocha}. Our proposal 
(\ref{BFM}) formally belongs to the same class and 
some details concerning its relation to the Dicke effect 
have been already in pointed out in the review paper 
\cite{Brandes-04}.

For the present context it is important to emphasize that 
the spectral function $\rho_{d,\sigma}(\omega\!=\!0)$ is 
temperature-independent. In order to clarify this property 
let us consider the bare Green's function (\ref{GF_at}) 
which at low energies $\omega \rightarrow 0$ simplifies to 
$G_{d}^{V_{{\bf k}\beta}\!=\!0}(\omega) \simeq {\cal{Z}}/
\omega$ (we refer here to the symmetric case $E_{d}\!=\!0$). 
Using the Ansatz (\ref{GF}) we obtain ${\cal{G}}_{d}(\omega)
\simeq \frac{\cal{Z}}{\omega + i{\cal{Z}}(\Gamma_L+\Gamma_R)
/2}$ which yields the following fixed value of the spectral 
function $\rho_{d,\sigma}(\omega) \simeq \frac{1}{\pi} \; 
\frac{2}{\Gamma_L+\Gamma_R}$. Furthermore we notice that 
effective broadening of a central peak depends on the 
spectral weight ${\cal{Z}}$, hence comes its temperature 
dependence which is illustrated in figure \ref{density}. Obviously, 
in a weak interaction regime $|g| \ll \Gamma_{\beta}$ the 
three peaks overlap with one another therefore the fine 
structure related to the charge Kondo state fades away.

\section{Transport properties}

\begin{figure}
\epsfxsize=9cm
\centerline{\epsffile{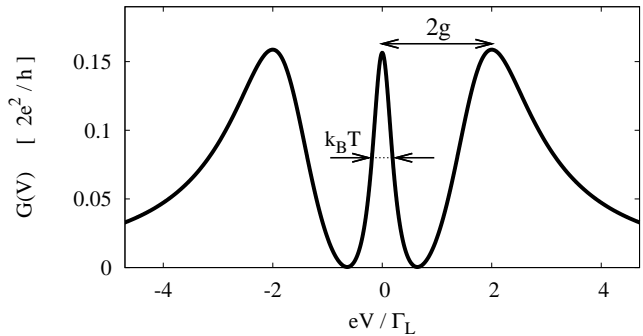}}
\caption{The differential conductance $G(V)$ as a function of 
bias voltage $V$ in the symmetric case for $k_{B}T\!=\!0.2
\Gamma_{L}$, $\Gamma_{R}\!=\!\Gamma_{L}$ and $g\!=\!\Gamma_{L}$. 
Width of the central (superradiance) peak is proportional 
to temperature whereas the zero bias value $G(0)$ remains 
constant (temperature-independent) \cite{Koch}.}
\label{conductance}
\end{figure}

External bias $V$ applied between the electrodes induces 
the charge and energy currents through the interface. We 
expect that behavior presented in figure \ref{density} is 
going to have an impact on the low-temperature transport 
properties. To determine the charge current $I_{\beta}(V)=
-e\frac{d}{dt}\langle \sum_{{\bf k}\sigma} \hat{c}^{+}_{
{\bf k} \beta \sigma} \hat{c}_{{\bf k} \beta \sigma} 
\rangle$ one has to use either the nonequilibrium Keldysh 
or Kadanoff-Baym formalism \cite{books}. Standard calculations 
\cite{Meir-93} yield the steady electron current $I(V)\!
=\!I_{L}(V)\!=\!-I_{R}(V)$ given by the Landauer-type 
formula
\begin{eqnarray}
I(V) =  \frac{2e}{h} \int_{-\infty }^{\infty } d\omega 
[f(\omega-\mu_L)\!-\!f(\omega-\mu_R) ]\; T(\omega), 
\label{I}
\end{eqnarray}
where $\mu_L\!-\!\mu_R=eV$ and $f(\omega)=\left[ 1+\mbox
{exp}(\omega)/k_{B}T \right]^{-1}$. Correlation effects 
show up within such approach through the transmission 
coefficient $T(\omega) =  \frac{\Gamma_{L} \Gamma_{R}}
{\Gamma_{L}+\Gamma_{R}}\; \sum_{\sigma} \rho_{d,\sigma}
(\omega)$. This treatment is valid for the wide-band limit 
$\Gamma_{\beta} \ll D$  and it can be derived, for 
instance from the equation of motion method \cite{Ng-96}.

We have explored numerically the differential conductance 
$G(V)\!=\!\frac{d}{dV}I(V)$ for a number of temperatures.
Figure \ref{conductance} shows a representative result 
obtained for $k_{B}T=0.2\Gamma_{L}$ where a middle peak
corresponds to the characteristic charge Kondo feature. 
Its width is proportional to temperature, whereas  
the zero bias conductance $G(0)\!=\!\frac{2e^{2}}{h} 
\int_{-\infty}^{\infty} d\omega \left[ -\;\frac{d 
f(\omega)}{d\omega} \right]\; T(\omega)$ at low 
temperatures approaches the asymptotic value $\lim_{T 
\rightarrow 0} G(0)=\frac{2e^{2}}{h} \; \frac{\Gamma_{L} 
\Gamma_{R}}{\Gamma_{L}+\Gamma_{R}}\; \sum_{\sigma}\rho_{d,
\sigma}(0)$. In the previous study \cite{Koch} of the charge 
Kondo physics such feature has been predicted for the zero-bias 
conductance as a function of the gate voltage. From our study
of the two-component model (\ref{BFM}) we find the same 
behavior but this issue will be discussed separately. 

\section{Summary}
We have studied the charge transport through a correlated 
quantum dot with an effective negative value of the Coulomb 
interaction $U<0$. To account for the quantum fluctuations 
between the empty and doubly occupied states we have 
introduced the phenomenological model where QD dot is coupled 
to an additional electron pair buffer via the Andreev-type 
interaction. Focusing on the symmetric case, when the Kondo 
effect arises in the pseudospin charge channel
\cite{Coleman-91}, we have examined its influence on the 
spectral function and the differential conductance. For 
both these quantities we have found the narrow peak whose 
height is constant whereas its width is proportional to
temperature. This resembles the properties of the Dicke 
effect \cite{Brandes-04} which has been recently considered 
for a configuration composed of the three vertically coupled 
quantum dots \cite{Trocha}. It might be of some interest 
to proceed with an analysis of here proposed model outside 
the symmetric case and adopt some more sophisticated 
selfconsistent treatment for the Andreev-like interaction 
on the molecular quantum dot.

\vspace{0.5cm}
This work is partly supported by the Polish Ministry of Science 
and Education under the grants NN202187833 and NN202373333.

\end{document}